\documentclass[11pt,dvips]{article}
cm \voffset = -2truecm

\usepackage{graphicx}
\usepackage{amssymb}
\usepackage{latexsym}

\begin{document}
\begin{titlepage}
\setcounter{page}{1}
\renewcommand{\thefootnote}{\fnsymbol{footnote}}

\vspace{5mm}
\begin{center}

 {\large \bf Generalized Grassmann variables for quantum kit ($k$-level) systems and  Barut-Girardello coherent states
             for $su(r+1)$ algebras}

\vspace{1.5cm}

{\bf M. Daoud}$^{a,b}${\footnote { email: {\sf m$_{-}$daoud@hotmail.com}}}  and  {\bf L. Gouba}$^{b}${\footnote { email: {\sf
lgouba@ictp.it}}}
\vspace{0.5cm}\\
$^{a}${\it Department of Physics , Faculty of Sciences Ain Chock,\\ University Hassan II,
Casablanca ,
Morocco}\\[1em]

$^{b}${\it Abdus Salam  International Centre for Theoretical Physics,\\ Strada Costiera 11,
I - 34151 Trieste, Italy}\\[1em]

\vspace{1.5cm}

\begin{abstract}
This paper concerns the construction of $su(r+1)$ Barut--Girardello  coherent states in term of
generalized Grassmann variables. We first introduce a generalized Weyl-Heisenberg algebra ${\cal A}(r)$ ($r \geq 1$) generated by $r$ pairs of creation and annihilation operators.
This algebra provides a useful framework to describe qubit and qukit ($k$-level) systems. It includes the usual Weyl-Heisenberg and
$su(2)$ algebras. We investigate the corresponding
Fock representation space. The generalized Grassmann variables are introduced as variables spanning the Fock--Bargmann space
associated with the algebra ${\cal A}(r)$. The  Barut--Girardello coherent states for $su(r+1)$  algebras are
explicitly derived and their over--completion properties are discussed.
\end{abstract}
\end{center}
\end{titlepage}

\newpage

\section{Introduction}

The coherent states formalism  has been  widely used in several areas of quantum physics \cite{Glauber,Gilmore, Klauder1, Perelomov1, Ali1, Gazeau2}.
The coherent states for quantum harmonic oscillator were initially introduced by Schr\"odinger \cite{Shrodinger}  and  were extended to other exactly solvable quantum systems (see for instance \cite{Gazeau1,AEG,Antoine,0411210,El Kinani1,El Kinani2,Daoud-Hussin} and references therein). The generalized
coherent states associated with  any irreducible unitary
representation of any Lie group were also investigated \cite{Gilmore,Glauber, Klauder1, Perelomov1, Ali1, Gazeau2}. The construction of coherent states has been proceeded  along
three (not equivalent in general) lines
 (see for instance \cite{Zhang}). The first line, due to Perelomov, generates the set of generalized coherent states
 by the action  of an unitary displacement operator
on a reference state of a group representation Hilbert space. The representation space might be finite or infinite dimensional  \cite{Perelomov1,Perelomov2}. The second
line, proposed by  Barut and Girardello, defines the
 generalized coherent states  as the eigenstates of the lowering generators \cite{Barut}. This approach applies only in the case of non-compact groups with
 infinite dimensional representation spaces. The third line is based
 on the  minimization of the Robertson-Schr\"odinger
uncertainty relations for the hermitian generators of the  group symmetry \cite{Shrodinger,Shrodinger1, Robertson} (see also \cite{Aragone2, Aragone1}).\\

The Perelomov and minimum Robertson-Schr\"odinger uncertainty approaches apply
for Lie symmetries with infinite as well finite  dimensional representation spaces. However, the
Barut--Girardello approach can be employed only
for non-compact Lie groups ($SU(1,1)$ for instance)  \cite{Barut}. Recently, the construction of $SU(2)$ coherent states of
Barut--Girardello type has been considered in \cite{DK} and it has been shown that the eigenstates of the lowering
angular momentum operator exist if the eigenvalues are no longer complex numbers
but are  variables generalizing the  usual Grassmann variables. One of the purposes of the present work is to continue  the
study reported in \cite{DK} in order to derive the  Barut--Girardello coherent states for $su(r+1)$ algebras $(r = 1, 2, \cdots)$
using the formalism of generalized Grassmann variables. We note that  the mathematical structures and the properties  of this  kind of variables  were
investigated in the context of non commutative geometry in several works (see for instance \cite{Majid,Kerner,Filippov,Isaev,Cugliandolo,Ilinski,Mansour1,Mansour2} ). They were employed to formulate the coherent states of quantum systems with finite
dimensional Hilbert space such as deformed harmonic oscillators \cite{DHKrusse,Daoud,Cabra,Maleki} and pseudo-Hermitian quantum systems \cite{Trifonov,Cherbal,Najabashi}.\\

The essential  algebraic tool of this paper is the generalized Weyl-Heisenberg algebra ${\cal A}(r)$ $(r \geq 1)$ generated by $r$ creation, $r$ annihilation and $r$ number
operators.  This algebra covers  the generalized Weyl-Heisenberg algebras  ${\cal A}(1)$ and ${\cal A}(2)$ introduced in \cite{daoud-kibler1,daoud-kibler2,daoud-kibler3}
and provides the appropriate algebraic tools to describe finite quantum systems such as qubits and qudits. Furthermore, the generalized Grassmann variables emerge naturally in the analytical representations
of these generalized oscillator algebras and provide us with the adequate ingredient to construct the  Barut--Girardello
 coherent states for $su(r+1)$ algebras. It must be emphasized that the idea of this work originates from the
recent developments discussed in \cite{DK} to determine  the Barut--Girardello coherent states for $su(2)$ algebra. \\

This paper is organized as follows. The first section introduces the generalized  Weyl-Heisenberg algebra ${\cal A}(r)$.
 The
corresponding finite dimensional Fock representations are also presented. In section 3, we develop the relation between the
Weyl-Heisenberg algebra ${\cal A}(1)$ and qukit systems ($k$-level systems). We also discuss the correspondence between
Dicke states and the Fock vectors generating  the representation space for the algebra  ${\cal A}(1)$. This
correspondence is useful to provide the realization of the generalized Grassmann variables in terms of the
usual Grassmann variables and subsequently to define the generalized Grassmann derivative and integration. Section
4 deals with the Fock--Bargmann representation of the algebra ${\cal A}(r)$. In this representation the creation operators
acts as  multiplication by a generalized Grassmann
variables. In section 5, we give  the $su(r+1)$ coherent states of Barut--Girardello labelled by $r$ complex variable and a generalized Grassmann
variable. The over-completion
property of the obtained sets of coherent states is examined.
Concluding remarks close this paper.

\section{Generalized Weyl-Heisenberg algebras}

\subsection{The algebra ${\cal A}(r)$ }

We begin by introducing the generalized Weyl-Heisenberg algebra ${\cal A}(r)$. This algebra is generated by $3r$ operators
 $a_i^-$, $a_i^+$ and $N_i$~($i=1,2, \cdots, r$).  They satisfy the commutation relations
			\begin{eqnarray}
[a_i^- , a^+_i] = k ~\mathbb{I} -  \bigg(\sum_{j=1}^{r}N_j + N_i\bigg), \quad
[N_i , a_j^{\pm}] = {\pm} \delta_{i,j} a_i^{\pm}, \quad i,j = 1,2, \cdots, r,
			\label{com1}
			\end{eqnarray}
and
			\begin{eqnarray}
[a_i^{\pm} , a_j^{\pm}] = 0, \quad i \neq j,
 			\label{commutation2}
			\end{eqnarray}
complemented by the triple relations
			\begin{eqnarray}
[a_i^{\pm} , [a_i^{\pm} , a_j^{\mp}]] = 0, \quad  i \neq j.
			\label{commutation3}
			\end{eqnarray}
In Eq.~(\ref{com1}), $\mathbb{I}$ denotes the identity operator
and $k \in \mathbb{N}^*$. This algebra extends  the generalized Weyl-Heisenberg algebras ${\cal A}(1)$ and ${\cal A}(2)$
introduced in \cite{daoud-kibler2,daoud-kibler3}. Indeed, for $r=1$,  the algebra  ${\cal A}(1)$ is  spanned by the three linear operators $a_-$, $a_+$ and $N$ satisfying the
following relations
\begin{equation}\label{A1}
  [a_-, a_+] = k ~\mathbb{I} - 2 N  \qquad [N, a_{\pm}] = \pm a_{\pm}.
\end{equation}
For $r=2$,  the algebra ${\cal A}(2)$  is generated by six linear
operators $a_i^-$, $a_i^+$ and $N_i$ ($i = 1, 2$) satisfying the structure  relations
			\begin{eqnarray}
[a_i^- , a^+_i] = k ~\mathbb{I} - (N_1 + N_2 + N_i), \quad
[N_i , a_j^{\pm}] = {\pm} \delta_{i,j} a_i^{\pm}, \quad i,j = 1,2
			\label{commutation1r2}
			\end{eqnarray}
and
			\begin{eqnarray}
[a_i^{\pm} , a_j^{\pm}] = 0, \quad i \neq j,\qquad [a_i^{\pm} , [a_i^{\pm} , a_j^{\mp}]] = 0, \quad i \neq j.
 			\label{commutation2r2}
			\end{eqnarray}
Remark that the ${\cal A}(r)$ algebra is similar  to the bosonic algebra introduced by Palev \cite{Palev1} which describes  a collection
of identical  particles  obeying  the so-called $A_r$--statistics. This
algebraic description was further investigated  from the microscopic point of view by Palev and Van der
Jeugt \cite{Palev2}. The creation and annihilation of particles
obeying  the $A_r$--statistics are identified with the $2r$ Weyl generators  of the  $su(r+1)$ Lie algebra \cite{Palev1,Palev2}. In this sense, it is interesting to note
that  the  $su(r+1)$ generators  can  be realized in terms of the creation and the annihilation
operators of  the generalized Weyl-Heisenberg algebra ${\cal A}(r)$ as follows
\begin{eqnarray}
& & E_{+\alpha} =  a_{\alpha}^+, \quad
		E_{-\alpha} =  a_{\alpha}^-, \quad \alpha = 1,2,\cdots, r \nonumber \\
& & H_i = \frac{1}{2} \bigg( k ~\mathbb{I} -  (\sum_{j=1}^{r}N_j + N_i)\bigg), \quad i = 1,2,\cdots, r. \nonumber
\end{eqnarray}
The creation and annihilation operators coincide with the Weyl generators $E_{\pm\alpha}$. The Cartan generators $H_i$ are
expressed in terms of the number operators $N_i$ ($i=1, 2, \cdots, r$). The  remaining $r^2 - r$ generators of  $su(r+1)$  algebra
are realized as the commutators between the annihilation and creation operators as
\begin{eqnarray}
E_{+\alpha, -\beta}= [a_{\alpha}^+ , a_{\beta}^-], \quad E_{+\beta, -\alpha}= [a_{\beta}^+ , a_{\alpha}^-] \quad (\alpha < \beta, \quad \alpha , \beta = 1,2,\cdots, r).
\label{lesdeuxa3}
\end{eqnarray}
We stress that the algebra ${\cal A}(r)$  is defined by means of $3r$ generators, satisfying commutation relations (\ref{com1}),(\ref{commutation2}) and the triple commutation
relations (\ref{commutation3}), rather than the usual $r(r+2)$ generators
for the $su(r+1)$ Lie algebra.

\subsection{The Fock representation}
 Let us denote the Hilbert-Fock space of the algebra ${\cal A}(r)$ by ${\cal F}$. It is defined by
\begin{equation}
{\cal F} =  \bigoplus_{n=0}^{\infty} {\cal H}^n,
\end{equation}
where ${\cal H}^n
\equiv \{ |n_1, n_2,\cdots , n_r\rangle\ , n_i \in \mathbb{N},
\sum_{i=1}^{r}n_i = n > 0\}$ and ${\cal H}^0 \equiv \mathbb{C}$.  The Fock states $|n_1, n_2,\cdots , n_r\rangle$ are
the eigenvectors of the operators number $N_i$:
$$N_i |n_1, n_2,\cdots , n_r\rangle = n_i |n_1, n_2,\cdots , n_r\rangle.$$
The action of the creation and annihilation operators $a_i^{\pm}$, on ${\cal F}$, are defined by
\begin{equation}
a_i^{\pm} |n_1,\cdots, n_i,\cdots , n_r\rangle\ = \sqrt{F_i
(n_1,\cdots ,n_i \pm 1,\cdots, n_r)}|n_1,\cdots, n_i \pm 1,\cdots
, n_r\rangle\
\end{equation}
 where the structure functions $F_i$ can  be determined by employing the structure relations (\ref{com1}),(\ref{commutation2}) and  (\ref{commutation3}). They should be  non-negatives so that all states are well defined.
 We assume that $|0, 0,\cdots , 0\rangle$ is the vacuum from which the states $|n_1,\cdots, n_i,\cdots , n_r\rangle$ are
 generated by repeated applications of the raising operators $a_i^+$. The condition $a_i^- |0, 0,\cdots , 0\rangle\ = 0$ implies that the functions
 $F_i(n_1,\cdots ,n_i ,\cdots, n_r)$ satisfy
 \begin{equation}
F_i (n_1,\cdots ,0,\cdots, n_r) = 0
\end{equation}
for any mode $i$ ($i=1, 2,\cdots, r$). Considering the commutation rules (\ref{com1}), one gets
the following recurrence relations
\begin{equation}
F_i (n_1,\cdots ,n_i+1,\cdots, n_r) - F_i (n_1,\cdots ,n_i+1,\cdots, n_r) = k - ( n_1 + n_2 + \cdots + n_r ) - n_i,
\end{equation}
from which one obtains
\begin{equation}\label{F-general}
F_i (n_1,\cdots ,n_i,\cdots, n_r) = n_i \big(k+1-(n_1+\cdots +n_i+\cdots+ n_r)\big).
\end{equation}
The
actions of the raising and lowering operators on the Hilbert-Fock space ${\cal F}$ are thus given by
\begin{equation}\label{actionaimoins-r}
a_i^{-} |n_1,\cdots, n_i,\cdots , n_r\rangle\ = \sqrt{n_i (k+1 -
(n_1+n_2+\cdots+n_r))}|n_1,\cdots, n_i-1,\cdots , n_r\rangle,
\end{equation}
\begin{equation}\label{actionaimoins+r}
a_i^{+} |n_1,\cdots, n_i,\cdots , n_r\rangle\ = \sqrt{(n_i+1) (k+1
-(n_1+n_2+\cdots+n_r+1))}|n_1,\cdots, n_i+1,\cdots , n_r\rangle.
\end{equation}
The positivity condition  of the structure functions $F_i$  given by
\begin{equation}
k+1 -(n_1+n_2+\cdots+n_r) > 0.
\end{equation}
determines the  dimension of the irreducible representation space ${\cal F}$. Indeed,
 there exists a finite number of  states satisfying this
condition and the Fock space dimension is  given  by $\frac{(k+r)!}{k!r!}$. We note that the operators
defined by (\ref{lesdeuxa3}) act in this representation as
\begin{equation}\label{ineqj+}
[a_i^{+}, a_j^-]~ |n_1,\cdots, n_i,\cdots,n_j ,\cdots, n_r\rangle\ = \sqrt{n_j(n_i+1))}~|n_1,\cdots, n_i+1,\cdots, n_j-1,\cdots , n_r\rangle,
\end{equation}
 \begin{equation}\label{ineqj-}
[a_j^{+}, a_i^-]~ |n_1,\cdots, n_i,\cdots,n_j ,\cdots, n_r\rangle\ = \sqrt{n_i(n_j+1))}~|n_1,\cdots, n_i-1,\cdots, n_j+1,\cdots , n_r\rangle.
\end{equation}
In the situation where $k$ is large,  the algebra ${\cal A}(r)$ reduces to $r$ commuting copies of the usual harmonic oscillator algebra.  Indeed, using the
equations (\ref{actionaimoins-r}), (\ref{actionaimoins+r}), (\ref{ineqj+}) and (\ref{ineqj-}) one has
$$ \bigg[\frac{a_i^{-}}{\sqrt{k}} , \frac{a_i^{+}}{\sqrt{k}} \bigg] \sim \mathbb{I}  \quad {\rm and} \quad   \bigg[\frac{a_i^{-}}{\sqrt{k}} , \frac{a_j^{+}}{\sqrt{k}} \bigg] \to 0 \quad {\rm for} \quad i\neq j , $$
which describes a $r$-dimensional quantum harmonic oscillator.

\section{Qukits and generalized Weyl-Heisenberg algebra}

Dealing with bosonic and fermionic many particles states is simplified  by considering the
algebraic structures of the corresponding raising and lowering operators. For bosons the creation and annihilation operators satisfy the commutations relations
\begin{equation}\label{CR}
 [ b_i^-  , b_j^+ ] = \delta_{ij} \mathbb{I}, \qquad [  b_i^-  , b_j^- ] = [  b_i^+  , b_j^+ ] = 0,
\end{equation}
where the unit operator $\mathbb{I}$ commute with the creation and annihilation operators $b_i^+$ and $b_i^-$.  On the hand,
fermions are specified by the following anti-commutation relations
\begin{equation}\label{ACR}
\{  f_i^-  , f_j^+  \} = \delta_{ij}\mathbb{I}, \qquad \{  f_i^+  , f_j^+  \} = \{  f_i^-  , f_j^-  \} = 0.
\end{equation}
The properties of Fock states follow from the commutation and anti-commutation relations which imposes only  one particle in each state
for fermions (two dimensions) and multiple particles for bosons (infinite dimension). Following Wu and Vidal \cite{Wu-Lidar} there is a crucial difference between fermions and qubits (two level systems). In fact, a
qubit  is a vector in a two dimensional Hilbert space like fermions and the  Hilbert space of a multi-qubit system  has a tensor product structure like bosons. In this respect,
the raising and lowering operators commutation rules for qubits  are neither specified by  relations of bosonic type (\ref{CR}) nor of fermionic type (\ref{ACR}).

\subsection{Qubit algebra from generalized Weyl-Heisenberg algebra}

The qubits appear like objects which  exhibits both bosonic and fermionic properties so that they cannot be described by Fermi-like or Bose-like operators.
An alternative way for the algebraic description of qukits ($(k+1)$-level quantum systems) is possible by resorting the formalism of generalized Weyl-Heisenberg algebras.
We denote by $\vert - \rangle$ the ground state and $\vert + \rangle$ the excited state of a two-level system (qubit) so that the lowering, raising and number operators are defined by
\begin{equation}\label{action-q}
q^- = \vert - \rangle \langle + \vert , \qquad     q^+ = \vert + \rangle \langle - \vert ,       \qquad     N_q = \vert - \rangle \langle + \vert.
\end{equation}
They satisfy the  commutation relations
\begin{equation}\label{kappamoinsun}
[q^- , q^+] = \mathbb{I}_2 - 2 N_q, \qquad [N_q , q^+] = - q^+, \qquad [N_q , q^-] = + q^-,
\end{equation}
where  $\mathbb{I}_2$ is the $2\times 2$ identity matrix. In this scheme, the qubit is described by the modified bosonic algebra (\ref{kappamoinsun}) and the creation and the annihilation
operators satisfy the nilpotency condition
\begin{equation}\label{q-nilp}
(q^+)^2 = (q^-)^2 = 0
\end{equation}
like Fermi operators. This representation  turns out to be a particular case of the finite dimensional representations of
the generalized Weyl-Heisenberg algebra ${\cal A}(1)$  (\ref{A1}) corresponding to the situation where $r=1$  with $k=1$. We note that
 the commutation relations (\ref{kappamoinsun})  coincide with ones defining the algebra introduced  in \cite{andrzej}
to provide  an alternative algebraic description of  qubits instead of the parafermionic formulation considered in \cite{Wu-Lidar}.

\subsection{Qukit algebra and Dicke states}

To extend the above qudit description to qukits ($(k+1)$-dimensional quantum systems ($k \in \mathbb{N}^{\ast}$)), we consider a collection of $k$ copies of the algebra
(\ref{kappamoinsun}) generated by the raising and lowering operators $q^+_i$ and $q^-_i$,
the number operators $N_{q_i}$  and the unit operator $\mathbb{I}_2$ satisfying the relations
\begin{equation}\label{RC-qubits}
[q^-_i , q^+_j] = (\mathbb{I}_2 - 2 N_{q_i})~\delta_{ij}, \qquad [N_{q_i} , q^+_j] = - \delta_{ij}q^+_j, \qquad [N_{q_i} , q^-_j] = + \delta_{ij}q^-_j \qquad [  q_i^-  , q_j^- ] = [  q_i^+  , q_j^+ ] = 0.
\end{equation}
where $i = 1, 2, \cdots, k$. Let denote by  ${\cal H}_i= \{  \vert m_i \rangle, ~~ m_i = -,+  \}$ the Hilbert space for the qubit $i$. In view of the relations $[q^-_i , q^-_j] = 0$ for $i\neq j$,
the multi-qubit Hilbert space has the following  tensor product structure
$$ {\cal H}(k) = \bigotimes_{i=1}^{k}{\cal H}_i = \{  \vert m_1, m_2, \cdots, m_{k} \rangle, ~~ m_i = -,+  \},$$
like bosons. We define the
collective  lowering and raising operators
 in the Hilbert space  $ \mathcal{H}(k)$ as follows
\begin{equation}\label{green}
a^- = \sum_{i=1}^{k} q_i^-  \qquad \quad a^+ = \sum_{i=1}^{k} q_i^+
\end{equation}
in terms of the creation and annihilation operators $q_i^+$ and  $q_i^- $. Here and in the following the index $i$ refers to the system the
operator is acting on, e.g.
$$ q_i^{\pm} \equiv \mathbb{I}_2\otimes \cdots \mathbb{I}_2\otimes q_i^{\pm}\otimes \mathbb{I}_2\otimes \cdots\mathbb{I}_2.  $$
It is simple
to see that the state $\vert -, -, \cdots, -\rangle \equiv \vert 0 \rangle$
satisfies $a^- \vert  0 \rangle = 0$. Furthermore, using the commutation relations (\ref{RC-qubits}), one gets the  nilpotency relations
\begin{equation}\label{nilpo1}
(a^-)^{k+1} = 0 \qquad (a^+)^{k+1}  = 0
\end{equation}
which extends the Pauli exclusion principle for ordinary qubits (i.e., $k=1$) described by the conditions (\ref{q-nilp}).  The actions of the operators $a^-$ and $a^+$ on the Hilbert space $ \mathcal{H}(k)$
can be determined from the standard actions of the fermionic operators $q_i^-$ and $q_i^+$ (cf. equations (\ref{action-q})).  Using  a recursive procedure, one verifies that  repeated applications  of the raising operator $a^+$  on the vacuum
 $\vert -, -, \cdots, -\rangle \equiv \vert 0 \rangle$ gives
\begin{equation}\label{equ3}
(a^+)^n \vert 0 \rangle = \sqrt{\frac{n!k!}{(k-n)!}} ~\vert n \rangle
\end{equation}
where the vectors $\vert  n \rangle$ are the symmetric Dicke states with $n$ excitations $(n = 0, 1, 2, \cdots, k)$. They are defined by
\begin{equation}\label{dicke}
 \vert  n\rangle = \sqrt{\frac{n! (k-n)!}{k!}} \sum_{\sigma\varepsilon S_{k}} \vert \underbrace{-, -,\cdots,-}_{k-n},\underbrace{+,+,\cdots,+}_n \rangle
\end{equation}
where $S_{k}$ is the permutation group of $k$ objets.  The Dicke states generate an orthonormal basis of the symmetric Hilbert subspace of dimension
$k+1$.  The explicit expressions of the actions of the ladder operators $a^{\pm}$ can be written using  the structure function  $F(n) = n(k+1-n)$ (\ref{F-general}) .
 The equation (\ref{equ3}) rewrites as
\begin{equation}\label{}
(a^+)^n \vert 0 \rangle = \sqrt{F(n)!}~ \vert  n \rangle
\end{equation}
where $F(n)! = F(n)F(n-1)\cdots F(1)$ and $F(0) = 1$. After some algebra, it is simple to verify that
\begin{equation}\label{actionsr=1}
a^+ \vert  n \rangle = \sqrt{F(n+1)}~ \vert n+1 \rangle, \qquad a^- \vert n \rangle = \sqrt{F(n)}~ \vert n-1 \rangle
\end{equation}
and the action of the creation and annihilation operators on the vectors $\vert k, 0 \rangle$ and $\vert k, k \rangle$ gives
\begin{equation}\label{}
a^- \vert  0 \rangle = 0  \qquad a^+ \vert  k \rangle = 0.
\end{equation}
The number operator $N$ is defined as
\begin{equation}\label{action-A}
N \vert  n \rangle = n~ \vert n \rangle.
\end{equation}
The qukit operators $a^+$, $a^-$ and $N$ satisfy the commutation rules
\begin{equation}\label{qudit}
[ a^+ , a^- ] = k \mathbb{I} - 2N , \qquad   [a^+ , N] =  a^+, \qquad   [a^- , N ] =  - a^-,
\end{equation}
which reflects that the algebra ${\cal A}(1)$ can be realized in terms of an ensemble of identical qubits.
Using the commutation relation $\left[ q^+_i, q^-_j \right ] = 0$ for $i\neq j$, it is  simple to verify that
$$ \left[ a^+, a^- \right ] = \sum _{i,j} \left[ q^+_i, q^-_j \right ] = \sum _{i} \left[ q^+_i, q^-_i \right ],$$
 and the operator $N$ can be expressed as
$$N = \sum _{i=1}^{k}N_{q_i} $$
 where $N_{q_i}$ is the single qubit number operator ($N_{q_i} \vert -\rangle_i = 0 $ and $N_{q_i} \vert +\rangle_i = \vert +\rangle_i$). It is remarkable that the creation and annihilation
 operator $a^+$ and $a^-$ close the following trilinear relation commutation
 $$[a^-, [ a^+ , a^- ]] = 2 a^-, \qquad  [a^+, [ a^+ , a^- ]] = -2 a^+ $$
characterizing a parafermion \cite{Palev1}. Note also that the definition (\ref{green}) is similar to Green decomposition in the
construction of parafermions from ordinary fermions. Therefore, the operators $a^+$,  $a^-$ and $N$ satisfying the relations  (\ref{qudit})
provide a simple algebraic description of $(k+1)$-level quantum systems (qukit).  This result shows the relevance of  generalized Weyl-Heisenberg algebras
in describing qukit systems. In particular, this realization expresses the Hilbert states of a qukit system in terms of Dicke states
of $k$ qubits. In this way, the global properties of the qukit system are encoded in an ensemble of $k$ identical  qubits. To close this section we note that
the algebraic realization  of qukit systems  provides a natural way to define the generalized Grassmann variables associated with the generalized Weyl-Heisenbeg
algebras possessing finite dimensional representation spaces.

\subsection{Generalized Grassmann variables}

We consider the algebra ${\cal G}$ generated by the identity $\mathbf{1}$ and $k$ commuting Grassmann variables $\theta_i$ ~ $(i = 1, 2, \cdots k)$ obeying the usual nilpotency conditions:
\begin{equation}\label{nilp-usuel}
\theta_i^2 = 0, \qquad [\theta_i, \theta_j] = 0.
\end{equation}
It is important to mention that in order to simplify our purpose,  we consider in this work a set of commuting Grassmann variables. We denote by $\bar{\theta}_i$ the complex conjugate of the element $\theta_i$. The algebra ${\cal G}$ is spanned by $2^k$ linearly
independent elements of the form $\theta_{i_1} \theta_{i_2}\cdots \theta_{i_n}$  with $i_1< i_2<\cdots<i_n$ for $ n = 0, 1,\cdots,k$. For $n=0$, the corresponding element is the identity.
The $\theta$-derivative $\partial_{\theta_i} = \frac{\partial}{\partial\theta_i}$ satisfies
\begin{equation}\label{partial-usuel}
\partial_{\theta_i}  \theta_j = \delta_{ij}, \quad \partial_i 1 = 0, \quad \partial_{\theta_i}  \partial_{\theta_j}  = \partial_{\theta_j} \partial_{\theta_i} .
\end{equation}
We define the generalized Grassmann variable as
\begin{equation}\label{ggv}
\eta= \sum_{i=1}^{k} \theta_i  \qquad \quad \bar{\eta}= \sum_{i=1}^{k} \bar{\theta}_i
\end{equation}
in terms of the nilpotent variables  $\theta_i$ and  $\bar{\theta}_i$.
We define the following symmetric $\theta$-polynomials
\begin{equation}\label{sym-eta}
 e_n (\vec{\theta}) = \sum_{i_1< i_2<\cdots<i_n} \theta_{i_1} \theta_{i_2}\cdots \theta_{i_n}, ~~{\rm for}~~ n=1,2,\cdots,k ~~{\rm and}~~e_0(\vec{\theta})=1.
\end{equation}
where $\vec{\theta} = (\theta_{1}, \theta_{2},\cdots, \theta_{n}) $.
Explicitly, we have
$$  e_1 (\vec{\theta}) =  \theta,~~  e_2 (\vec{\theta}) = \sum_{i<j} \theta_i \theta_j,~~
e_3 (\vec{\theta}) = \sum_{i<j<l} \theta_i \theta_j \theta_l,\cdots, ~~ e_k (\vec{\theta}) =  \theta_1 \theta_2 \cdots \theta_{k}.$$
The $n$-th power of the variable $\eta$ (\ref{ggv}) writes in term of the symmetric $\theta$-polynomials (\ref{sym-eta}) as
$$\eta^n = n!  e_{n}(\vec{\theta}) ~~{\rm for}~~ n=1,2,\cdots,k $$
and  the nilpotency conditions (\ref{nilp-usuel}) for ordinary Grassmann numbers implie $$\eta^{k+1} = 0.$$
Furthermore, we define the $\eta$-derivative as follows
\begin{equation}\label{eta-derivative}
\frac{\partial}{\partial \eta}= \sum_{i=1}^{k} \frac{\partial}{\partial \theta_i},  \qquad \quad  \frac{\partial}{\partial \bar{\eta}}= \sum_{i=1}^{k}  \frac{\partial}{\partial \bar{\theta}_i}.
\end{equation}
Using the properties of $\theta_i$-derivatives (\ref{partial-usuel}), one shows
$$\partial_{\eta}^n = n!  g_{n} ~~{\rm for}~~ n=1,2,\cdots,k ~~{\rm and}~~\partial_{\eta}^{k+1} = 0$$
where the differential operator $g_n$ is given by
\begin{equation}
 g_n  = \sum_{i_1< i_2<\cdots<i_n} \partial_{\theta_{i_1}}  \partial_{\theta_{i_2}}\cdots  \partial_{\theta_{i_n}}, ~~{\rm for}~~ n=1,2,\cdots,k ~~{\rm and}~~g_0=1.
\end{equation}
Using the symmetric $\theta$-polynomials (\ref{sym-eta}),  we define the Grassmann analogue of  Dicke  states (\ref{dicke}) as
\begin{equation}
D_{n}(\vec{\theta}) = \sqrt{\frac{n! (k-n)!}{k!}}  e_{n}(\vec{\theta}).
\end{equation}
It is interesting to note that the $n$-th power of the generalized Grassmann variable (\ref{ggv}) express as
\begin{equation}\label{power-eta}
\eta^n = \sqrt{\frac{n!k!}{(k-n)!}} D_{n}(\vec{\theta}) = n!  e_{n}(\vec{\theta})
\end{equation}
 in term of the functions $D_{n}(\vec{\theta})$ from which one shows
\begin{equation}
\eta D_{n}(\vec{\theta}) = \sqrt{(n+1)(k-n)} ~D_{n+1}(\vec{\theta}).
\end{equation}
This relation is similar to the action of the creation operator given by (\ref{actionsr=1}).  To obtain  the
derivative of the functions $D_{n}(\vec{\theta})$, we employ first the definition (\ref{eta-derivative}) to get the
derivative of $\theta$-polynomials (\ref{sym-eta}):
\begin{equation}
\frac{\partial e_{n}(\vec{\theta}) }{\partial \eta} = (k-n) ~e_{n-1}(\vec{\theta})  ~~{\rm for}~~ n=1,2,\cdots,k ~~{\rm and}~~\frac{\partial e_{0}(\vec{\theta}) }{\partial \eta} = 0,
\end{equation}
from which one gets
\begin{equation}
\frac{\partial{D_{n}(\vec{\theta})}}{\partial\eta} =  \sqrt{n(k+1-n)} ~D_{n-1}(\vec{\theta}).
\end{equation}
This result is similar to the action of the annihilation operation of the qukit algebra ${\cal A}(1)$ given by (\ref{actionsr=1}).
In this scheme, the $\eta$-integral can be derived from the Berezin integral of Grassmann variables given by
\begin{eqnarray}
	\int {\theta}_i  d{\theta_j} = \delta_{ij} \quad \int  d{\theta_i} = 0.
	\label{Berezin}
	\end{eqnarray}
Using the Berezin integration formula, it is simple to verify that the symmetric $\theta$-polynomials (\ref{sym-eta}) satisfy
$$\int  e_n (\vec{\theta})  d{\eta}
	 =0 \quad (n = 0, 1, \cdots, k-2) ~~~{\rm and} \int  e_{k} (\vec{\theta})  d{\eta}
	 = 1$$
where $d{\eta} = d{\theta_1}d{\theta_2}\cdots d{\theta_{k}}$. This gives  the following  $\eta$-integral formulas
\begin{eqnarray}\label{Berezin-gen1}
\int  \eta^n ~ d{\eta}  =0 \quad (n = 0, 1, \cdots, k-1) ~~~{\rm and} \int  \eta^{k} ~ d{\eta}   = k!.
	\end{eqnarray}
The usual Berezin integration for ordinary Grassmann
variables is recovered for $k=1$.  Similarly, for the conjugate generalized Grassmann variables we have the following integration rules
	\begin{eqnarray}\label{Berezin-gen2}
	\int {\bar {\eta}}^n d{\bar {\eta}} = 0 \quad (n = 0, 1, \cdots, k-1), \quad \int {\bar {\eta}}^{k} d{\bar {\eta}} = k!.
	\end{eqnarray}
The generalized integration formulas (\ref{Berezin-gen1}) and (\ref{Berezin-gen2}) are
of especially important   in deriving the over-completion property of the Barut--Girardello coherent of $su(r+1)$. This issue is discussed in Section 5.

\section{Fock--Bargmann realization of generalized Weyl-Heisenberg algebra}

In the Fock--Bargmann representation of the usual  Heisenberg-Weyl algebra, the
 creation and annihilation operators are respectively realized as multiplication and derivation
 with respect a complex variable \cite{Fock,Bargmann}. This representation is widely used in several problems of
 quantum physics and mathematics \cite{Perelomov1}. Thus, it is natural to determine the Fock--Bargmann spaces for algebras
 of ${\cal A}(r)$  type. In this sense,   the main goal of  this section concerns the characterization of any vector state in
 the Hilbert space of the generalized algebra ${\cal A}(r)$ by an analytic function expressed in terms of generalized Grassmann variables.

\subsection{The analytical representation of the generalized algebra ${\cal A}(1)$}

We shall first discuss the analytical realization of the algebra  ${\cal A}(1)$
 in which creation operator $a^+$ acts as multiplications by the variable $\eta$.
We realize the Fock space basis and the creation operation as
	\begin{equation}
	\vert n \rangle \longrightarrow f_{n}(\eta) =  c_{n} {{\eta}}^{n},  ~\qquad
	a^+ \longrightarrow {{\eta}}.
	\label{correspondancer1BG}
	\end{equation}
The nilpotency relation (\ref{nilpo1}) implies that the variable $\eta$  satisfies
\begin{eqnarray}\label{muli-nilp}
\eta^{k+1} = 0
\end{eqnarray}
From the correspondence (\ref{correspondancer1BG}), we write
\begin{equation}
	\vert 0 \rangle \longrightarrow  1
	\label{correspondance1BG}
	\end{equation}
such that  $c_{0}= 1$. Using the actions of the creation and annihilation operators (\ref{actionsr=1}), one verifies
\begin{equation}
	\vert n\rangle =  \sqrt{\frac{(k-n)!}{k!n!}}{({a^+})}^{n}  \vert 0 \rangle  \qquad n  \leq k,
	\label{correspondanceetats}
	\end{equation}
and using (\ref{correspondancer1BG}),  one gets $c_n = \sqrt{(k-n)!/k!n!}$. therefore,  the analytical functions $f_{n}(\eta)$ read as
\begin{equation}
f_{n}(\eta)  =  \sqrt{\frac{(k-n)!}{k!n!}}  {{\eta}}^{n}.
\label{correspondancefonvtions}
\end{equation}
It is simple to check that the multiplication and the derivative of the functions $f_{n}(\eta)$ with respect to the variable $\eta$  leads to
\begin{equation}
\eta ~ f_{n}(\eta)  =  \sqrt{(n+1)(k-n)}   f_{n-1}(\eta), \quad \frac{\partial}{\partial\eta} ~ f_{n}(\eta)  =  \sqrt{n(k+1-n)}   f_{n-1}(\eta).
\label{derivative}
\end{equation}
From the last equation, one verifies that the $\eta$-derivative satisfy the condition
\begin{equation}\label{der-nilp}
\bigg(\frac{\partial}{\partial\eta}\bigg)^{k+1}   = 0.
\label{derivative}
\end{equation}
For $k=1$, the qukit becomes a qubit system and the relations (\ref{muli-nilp}) and (\ref{der-nilp}) reduces to the
nilpotency conditions for the usual Grassmann variables.

\subsection{The analytical representation of the generalized algebra ${\cal A}(2)$}

The ${\cal A}(2)$ algebra  is spanned by two pairs of
creation and annihilation operators  $a_i^-$, $a_i^+$ with $i=1,2$ and two number operators $N_i$. They  satisfy the  structures  relations
(\ref{com1}), (\ref{commutation2}) and (\ref{commutation3}). The dimension of the Fock space
${\cal F}_{k} = \{ \vert n_1 , n_2 \rangle : n_1 \in \mathbb{N} , n_2 \in \mathbb{N}; ~n_1 + n_2 \leq k \}$
is  $d = (k+1)(k+2)/2$. The vectors $\vert n_1 , n_2 \rangle$ are  the eigenstates of the number operators $N_1$ and $N_2$
$(N_i \vert n_1, n_2 \rangle = n_i \vert n_1, n_2 \rangle, ~i=1,2)$.  From the equations (\ref{actionaimoins-r}) and (\ref{actionaimoins+r}), the raising and lowering operators  $a_1^{\pm}$ and $a_2^{\pm}$ act as
		\begin{eqnarray}
a_1^+ \vert n_1, n_2 \rangle  = \sqrt{(n_1+1)(k-n_1-n_2)} \vert n_1+1, n_2 \rangle, \quad
a_1^- \vert n_1, n_2\rangle = \sqrt{n_1(k+1-n_1-n_2)} \vert n_1-1,n_2\rangle
		\label{action1+-}
		\end{eqnarray}
and
		\begin{eqnarray}
a_2^+ \vert n_1, n_2 \rangle = \sqrt{(n_2+1)(k-n_1-n_2)}  \vert n_1,n_2+1 \rangle,\quad
a_2^- \vert n_1, n_2\rangle = \sqrt{n_1(k+1-n_1-n_2)} \vert n_1,n_2-1\rangle.
		\label{action2+-}
		\end{eqnarray}
Using the actions (\ref{action1+-}) and (\ref{action2+-}), one verifies that  the creation and annihilation operators satisfy the conditions
\begin{eqnarray}
(a_1^+)^{k+1-l}  (a_2^+)^{l} = 0, \quad (a_1^-)^{k+1-l}  (a_2^-)^{l} = 0, \quad {\rm for}\quad l = 0, 1,2, \cdots ,k+1.
\label{nilpo-hyb}
\end{eqnarray}
To discuss the analytical realization of the algebra ${\cal A}(2)$,
 in which creation operator $a_i^+$ acts as multiplications by the generalized Grassmann variables $\eta_i$ with $i=1,2$, we realize the Fock space basis as
	\begin{equation}
	\vert n_1, n_2 \rangle \longrightarrow f_{n_1,n_2}(\eta_1,  \eta_2) =  c_{n_1,n_2} {{\eta_1}}^{n_1} {{\eta_2}}^{n_2} \qquad
	a_i^+ \longrightarrow {{\eta_i}}.
	\label{correspondance1BG}
	\end{equation}
Using the relations  (\ref{nilpo-hyb}), the variables $\eta_1$ and $\eta_2$ satisfy
\begin{eqnarray}
(\eta_1)^{k+1-l}  (\eta_2)^{l} = 0, \qquad (\eta_1)^{k+1-l}  (\eta_2)^{l} = 0,
\end{eqnarray}
for $l =  0, 1,2, \cdots ,k+1$.
From the correspondence (\ref{correspondance1BG}), one can write
\begin{equation}
	\vert 0, 0 \rangle \longrightarrow  1
	\label{correspondance1BG}
	\end{equation}
where we set $c_{0,0}= 1$. Using the actions of the creation and annihilation operators (\ref{action1+-}) and (\ref{action2+-}), one verifies
\begin{equation}
	\vert n_1, n_2 \rangle =  \sqrt{\frac{(k-n_1-n_2)!}{k!n_1!n_2!}}{({a^+_1})}^{n_1} {({a^+_2})}^{n_2} \vert 0, 0 \rangle  \qquad n_1+n_2 \leq k,
	\label{correspondanceetats}
	\end{equation}
from which one gets the expressions of the analytical functions $f_{n_1,n_2}(\eta_1,  \eta_2)$  associated with the Fock states $\vert n_1, n_2 \rangle$
\begin{equation}
f_{n_1,n_2}(\eta_1,  \eta_2)  =  \sqrt{\frac{(k-n_1-n_2)!}{k!n_1!n_2!}}  {{\eta_1}}^{n_1} {{\eta_2}}^{n_2}.
\label{correspondancefonvtions}
\end{equation}
The derivative  with respect to the variables $\eta_1$ and $\eta_2$ gives
\begin{equation}
\frac{\partial}{\partial\eta_1} ~ f_{n_1,n_2}(\eta_1,  \eta_2)  =  \sqrt{n_1(k+1-(n_1+n_2))}   f_{n_1-1,n_2}(\eta_1,  \eta_2),
\end{equation}
\begin{equation}
\frac{\partial}{\partial\eta_2} ~ f_{n_1,n_2}(\eta_1,  \eta_2)  =  \sqrt{n_2(k+1-(n_1+n_2))}   f_{n_1,n_2-1}(\eta_1,  \eta_2).
\label{derivative}
\end{equation}
It follows that the derivatives satisfy the conditions
\begin{equation}
\bigg(\frac{\partial}{\partial\eta_1}\bigg)^{k+1-l} \bigg(\frac{\partial}{\partial\eta_2}\bigg)^{l}  = 0; \quad l = 0, 1, 2, \cdots, k+1,
\label{derivative}
\end{equation}
which gives the analytical analogue of the hybrid nilpotency relations satisfied by the annihilation operators $a_1^-$ and $a_2^-$ given by the equations (\ref{nilpo-hyb}).

\subsection{The analytical representation of the generalized algebra ${\cal A}(r)$}

The Fock--Bargmann representations corresponding to  ${\cal A}(1)$ and ${\cal A}(2)$ algebras can be easily extended to higher   ranks $r$.
Indeed,  the representation  space basis  of the algebra ${\cal A}(r)$  can be realized as
	\begin{equation}
	\vert n_1, n_2, \cdots, n_r \rangle \longrightarrow f_{n_1,n_2, \cdots, n_r }(\eta_1,  \eta_2, \cdots, \eta_r) =  c_{n_1,n_2,\cdots, n_r } {{\eta_1}}^{n_1} {{\eta_2}}^{n_2} \cdots  {{\eta_r}}^{n_r} \qquad
	a_i^+ \longrightarrow {{\eta_i}}.
	\label{correspondance1BG}
	\end{equation}
In particular, we assume that  the ground state is represented as
\begin{equation}
	\vert 0, 0, \cdots, 0 \rangle \longrightarrow  1,
	\end{equation}
by considering  $c_{0,0, \cdots, 0}= 1$. Using the actions of the creation and annihilation operators (\ref{actionaimoins-r}) and (\ref{actionaimoins+r}), one verifies
\begin{equation}
	\vert n_1, n_2, \cdots, n_r \rangle =  \sqrt{\frac{(k-n_1-n_2\cdots - n_r)!}{k!n_1!n_2!\cdots n_r!}}{{(a^+_1)}}^{n_1} {{(a^+_2)}}^{n_2} \cdots {{(a^+_r)}}^{n_r} \vert 0, 0,\cdots, 0 \rangle
\quad n_1+n_2+\cdots+n_r \leq k,
	\label{correspondanceetats}
	\end{equation}
and the function $f_{n_1,n_2, \cdots, n_r }(\eta_1,  \eta_2, \cdots, \eta_r)$, representing an arbitrary Fock vector $\vert n_1, n_2, \cdots, n_r \rangle $,  is given by
\begin{equation}
f_{n_1,n_2, \cdots, n_r }(\eta_1,  \eta_2, \cdots, \eta_r) =  \sqrt{\frac{(k-n_1-n_2\cdots - n_r)!}{k!n_1!n_2!\cdots n_r!}}  {{\eta_1}}^{n_1} {{\eta_2}}^{n_2}\cdots {{\eta_r}}^{n_r}.
\label{correspondancefonvtions}
\end{equation}
From the equation (\ref{correspondanceetats}), one can see that the creation operators satisfy the conditions
\begin{equation}
(a^+_1)^{l_1} (a^+_2)^{l_2} \cdots (a^+_r)^{l_r} = 0 \quad l_1 + l_2+\cdots l_r = k+1.
\end{equation}
Hence,  the variables $\eta_i$ satisfy the following generalized nilpotency conditions
\begin{eqnarray}
(\eta_1)^{l_1}  (\eta_2)^{l_2} \cdots  (\eta_r)^{l_r} = 0 \quad l_1 + l_2+\cdots l_r = k+1.
\end{eqnarray}
Using the derivative formula with respect a generalized Grassmann variable discussed in the previous section, one obtains the derivative of the functions $f_{n_1,n_2, \cdots, n_r }(\eta_1,  \eta_2, \cdots, \eta_r)$
\begin{equation}
\frac{\partial}{\partial\eta_i} ~ f_{n_1, \cdots,  n_i, \cdots, n_r}(\eta_1,\cdots,  \eta_i, \cdots, \eta_r)  =  \sqrt{n_i(k+1-(n_1+\cdots+n_i+\cdots+n_r))}   f_{n_1, \cdots,  n_i-1, \cdots, n_r}(\eta_1,\cdots,  \eta_i, \cdots. \eta_r)
\end{equation}
From this result, one has  the following nilpotency conditions
\begin{equation}
\bigg(\frac{\partial}{\partial\eta_1}\bigg)^{l_1} \bigg(\frac{\partial}{\partial\eta_2}\bigg)^{l_2} \cdots \bigg(\frac{\partial}{\partial\eta_r}\bigg)^{l_r}   = 0,\quad l_1 + l_2+\cdots l_r = k+1,
\label{derivative-r}
\end{equation}
which reduces to the conditions (\ref{der-nilp}) and (\ref{derivative}) for $r=1$ and $r=2$, respectively.

\section{ Barut-Girardello coherent states of $su(r+1)$ algebras}

\subsection{ The spin coherent states \`a la Barut-Girardello}
 The irreducible set  associated with a spin $j$ ($2j \in \mathbb{N}$) 
is spanned by the basis
   \begin{eqnarray}
B_{2j+1} = \{ |j , m \rangle : m = j, j-1, \ldots, -j  \},
   \end{eqnarray}
where  $|j , m \rangle$ is a common eigenvector of the Casimir
operator $j^2$ and of the Cartan operator $j_z$ of the Lie algebra $su(2)$:
          \begin{eqnarray}
          j^2 |j , m \rangle = j(j+1) |j , m \rangle, \quad
          j_z |j , m \rangle = m      |j , m \rangle.
          \label{jdeux}
          \end{eqnarray}
 The raising and lowering operators are given by
   \begin{eqnarray}
   j_+= \sum_{m = -j}^j {\sqrt{ (j+m+1)(j-m) }} |j , m +1 \rangle \langle j , m |, \quad  j_-= \sum_{m = -j}^j {\sqrt{ (j+m)(j-m+1) }} |j , m-1 \rangle \langle j , m |.
   \label{definition of h}
   \end{eqnarray}
They satisfy the structure relations
     \begin{eqnarray}
  \left[ j_z,j_{+} \right] = + j_{+},  \quad
  \left[ j_z,j_{-} \right] = - j_{-},  \quad
  \left[ j_+,j_- \right]   = 2j_z.
     \label{adL su2}
     \end{eqnarray}
In what follows, we make the identifications
$$ \vert j , m \rangle \longleftrightarrow \vert  n \rangle \qquad j+m \longleftrightarrow n, $$
so that the Hilbert space $B_{2j+1}$ is given by
$$  B_{2j+1} =  \{ |n \rangle : n = 0, 1, \ldots, 2j  \}.$$
To construct the $su(2)$ coherent states \`a la Barut-Girardello, one has to determine the eigenstates of the lowering operator $j_-$ by solving the following eigenvalue equation
     \begin{eqnarray}\label{su2-eigen}
 j_- \vert \lambda \rangle = \lambda \vert \lambda \rangle ~~ \qquad ~~ \vert \lambda \rangle  = \sum_{n = 0}^{2j} C_n \lambda^n \vert  n \rangle.
     \end{eqnarray}
 Using the action of $j_-$ (\ref{definition of h}), one gets the following recurrence relation
       \begin{eqnarray}\label{recu-rel}
C_{n+1}~ \sqrt{(n+1)(2j-n)} = C_n, \quad {\rm for} ~~~n = 0,1, \cdots, 2j-1,
     \end{eqnarray}
with the extremal condition
       \begin{eqnarray}\label{nilp}
C_{2j}~ \lambda^{2j+1} = 0.
     \end{eqnarray}
From the equations (\ref{recu-rel}) and (\ref{nilp}), it is clear that the eigenvalue equation admits a solution if the variable $\lambda$ satisfies
       \begin{eqnarray}\label{nilp1}
 \lambda^{2j+1} = 0.
     \end{eqnarray}
The eigenvalue  $\lambda$ can be written
       \begin{eqnarray}
 \lambda  =  \eta z,
     \end{eqnarray}
where $\eta$ is a generalized Grassmann variable of order $2j+1$ and $z$ is an arbitrary complex variable which commute with $\eta$.
The coefficients in the expansion of the states $\vert \eta \rangle$ (\ref{su2-eigen}) writes
       \begin{eqnarray}\label{coeff-cn}
 C_n = \sqrt{\frac{(2j-n)!}{n!(2j)!}} ~ C_0,
     \end{eqnarray}
where $C_0$ can be fixed from the normalization condition  of the states $\vert \lambda \rangle \equiv \vert \eta , z \rangle  $. As result one obtains
       \begin{eqnarray}\label{BG-Cs}
   \vert \eta , z \rangle = {\cal N}_2\sum_{n = 0}^{2j}  \sqrt{\frac{(2j-n)!}{n!(2j)!}} ~\eta^n z^n\vert  n \rangle,
     \end{eqnarray}
where the normalization factor is given by
$$ \vert {\cal N}_{2} \vert^{-2} = \sum_{n=0}^{2j} \frac{(2j-n)!}{n!(2j)!} \bar{\eta}^n \eta^n z^n \bar z^n.$$
It is remarkable that the states $\vert \eta, z \rangle$ reduce for $z=1$ to the so-called
$(2j+1)$-fermionic coherent states introduced in \cite{DK}. In addition,  they  constitute an over-complete set
	\begin{equation}\label{condition mesure222}
	\int \vert \eta , z \rangle d\mu (\eta ,\bar \eta, z, \bar z )
	\langle \eta , z  \vert = \sum_{n = 0}^{2j} \vert n \rangle \langle n \vert,
	\end{equation}
where  the measure $d\mu(\eta ,\bar \eta, z, \bar z )$ takes the form
\begin{equation}\label{condition mesure22}
d\mu(\eta ,\bar \eta, z, \bar z ) = \vert {\cal N}_2 \vert^{-2} \sigma_{2j}(\eta, \bar \eta) d\eta d\bar\eta~ dm( z, \bar z ),
\end{equation}
where the measure involving the complex variable $z$
$$
dm( z, \bar z ) =  e^{-\vert z \vert^2 }~\frac{d^2z}{\pi},
$$
coincides with the standard one with respect which the Glauber coherent states of the usual Weyl-Heisenberg (quantum harmonic oscillator)
satisfy the identity resolution property. Reporting (\ref{condition mesure22}) in (\ref{condition mesure222}) implies  that the function $\sigma_{2j}(\eta, \bar \eta)$ must verify the integral formula
	\begin{equation} \label{condition mesure2}
\int \sigma_{2j}(\eta, \bar \eta) d\eta d\bar\eta  \eta^{n} \bar\eta^{n} = \frac{(2j)!}{(2j-n)!}.
\end{equation}
Expanding the function $\sigma_{2j} (\eta ,\bar \eta )$ as
$$\sigma_{2j} (\eta ,\bar \eta ) = \sum_{n=0}^{2j} a_n \eta^{2j-n} \bar \eta^{2j-n}, $$
and using the generalized Berezin integration (\ref{Berezin-gen1}) and (\ref{Berezin-gen2}), one check that the integral equation (\ref{condition mesure2})  holds  for $a_n = \frac{1}{(2j)!(2j-n)!}$ and the explicit form of the function $\sigma_{2j} (\eta ,\bar \eta )$  is then given by
	\begin{equation}
\sigma_{2j}(\eta, \bar \eta) = \sum_{n=0}^{2j} \frac{1}{(2j)!(2j-n)!}\eta^{2j-n} \bar\eta^{2j-n}.
	\label{mesure grassmann}
	\end{equation}

\subsection{ The $su(3)$ coherent states \`a la Barut-Girardello}

Inspired by the Barut--Girardello coherent states for spin systems, we continue with the
 $SU(3)$ symmetry. The generators  of the $su(3)$ algebra are denoted   by
 $j_i^-$, $j_i^+$ and $j^0_i$ with $i=1,2$. They satisfy the following commutation relations
			\begin{eqnarray}
[j_i^+ , j^-_i] = 2 j^0_i, \quad
[j^0_i , j_j^{\pm}] = {\pm} \delta_{i,j} j_i^{\pm}, \quad i,j = 1,2, \quad [j_1^{\pm} , j_2^{\pm}] = 0,
			\label{commutation11}
			\end{eqnarray}
complemented by the triple relations
			\begin{eqnarray}
[j_1^{\pm} , [j_1^{\pm} , j_2^{\mp}]] = 0, \quad [j_2^{\pm} , [j_2^{\pm} , j_1^{\mp}]] = 0.
			\label{commutation33}
			\end{eqnarray}
 The remaining $su(3)$ operators are defined by
	\begin{eqnarray}
j_3^+ = [j_2^+ , j_1^-], \quad j_3^- = [j_1^+ , j_2^-].
\label{lesdeuxj3}
\end{eqnarray}
in terms of the  $j_i^-$, $j_i^+$ $(i=1,2)$ and the corresponding commutation rules are encoded in the  triple structure relations (\ref{commutation33}).
The dimension $d(s, t)$ of the irreducible representation $(s, t)$
of $SU(3)$ is given by
		\begin{eqnarray}
d(s, t) = \frac{1}{2}(s + 1)(t + 1)(s + t + 2 ),
\quad s \in \mathbb{N}, \quad t \in \mathbb{N}.
		\nonumber
		\end{eqnarray}
For  the irreducible representation $(0, k)$, or its adjoint $(k, 0)$, the dimension of representation space is given by
		\begin{eqnarray}
d = \frac{1}{2}(k+1)(k+2), \quad k \in \mathbb{N}^*.
		\label{dimensiond}
		\end{eqnarray}
The associated  Hilbert-Fock space is
 	\begin{eqnarray}
{\cal B}_{k}  = \{ \vert n_1 , n_2 \rangle :  n_1 + n_2 = 0, 1, \ldots, k \}
 	\nonumber
 	\end{eqnarray}
where the  vectors generating the orthonormal basis of ${\cal B}_{k}$ are the eigenstates of the the
number operators $N_1$ and $N_2$:
		\begin{eqnarray}
N_i \vert n_1, n_2 \rangle = n_i \vert n_1, n_2 \rangle, \quad i=1,2.
		\label{actionN}
		\end{eqnarray}
The action of the raising and lowering  operators $j_1^{\pm}$ and $j_2^{\pm}$ are given by
\begin{eqnarray}
j_1^+ \vert n_1, n_2 \rangle = \sqrt{(n_1+1)[ k -  (n_1+n_2)]}  \vert n_1+1, n_2 \rangle,~ j_1^- \vert n_1, n_2\rangle = \sqrt{n_1[ k +1 -  (n_1+n_2)]} \vert n_1-1,n_2\rangle,
  	\label{action1+}
  \end{eqnarray}
  and
\begin{eqnarray}
j_2^+ \vert n_1, n_2 \rangle = \sqrt{(n_2+1)[ k -  (n_1+n_2)]} \vert n_1,n_2+1 \rangle,~ j_2^- \vert n_1, n_2\rangle = \sqrt{n_2[ k +1 -  (n_1+n_2)]} \vert n_1,n_2-1\rangle,
		\label{action2+}
		\end{eqnarray}
The actions of the Cartan generators read as
\begin{eqnarray}
j^0_1\vert n_1, n_2\rangle = \bigg[n_1+\frac{1}{2}(n_2 - k)\bigg] \vert n_1,n_2\rangle,\quad
j^0_2\vert n_1, n_2\rangle =  \bigg[n_2+\frac{1}{2}(n_1 - k)\bigg]  \vert n_1,n_2\rangle.
		\label{actionh2}
		\end{eqnarray}
The actions of  the operators  $(j_3^+, j_3^-)$ defined by (\ref{lesdeuxj3})
in terms of the generators $(j_1^+, j_1^-)$ and $(j_2^+, j_2^-)$ can be determined  from (\ref{action1+})-(\ref{action2+}). One obtains
		\begin{eqnarray}
j_3^+ \vert n_1, n_2 \rangle = \sqrt{n_1(n_2 + 1)} \vert n_1-1 , n_2+1 \rangle, \qquad
j_3^- \vert n_1, n_2 \rangle = \sqrt{(n_1 + 1)n_2} \vert n_1+1 , n_2-1 \rangle.
		\end{eqnarray}
The $su(3)$ lowering operators  $j_1^-$ and $j_2^-$ commute and therefore can be diagonalized simultaneously. To find
the common set  of eigenstates we consider the solutions of
the following eigenvalue equations
		\begin{eqnarray}
j_1^-  \vert \lambda_1 , \lambda_2 \rangle = \lambda_1 \vert \lambda_1 , \lambda_2   \rangle, \quad j_2^-  \vert \lambda_1 , \lambda_2 \rangle = \lambda_2  \vert \lambda_1 , \lambda_2  \rangle.
		\label{Eqlambda}
		\end{eqnarray}
Expanding the state $\vert \lambda_1 , \lambda_2 \rangle $  as
		\begin{eqnarray}
\vert \lambda_1 , \lambda_2 \rangle   = \sum_{l=0}^{k} \sum_{n=0}^{k-l} C_{n,l} \lambda_1^n  \lambda_2^l  \vert n,l \rangle,
		\label{vectlambda}
		\end{eqnarray}
 the eigenvalues equation (\ref{Eqlambda}) of the first mode gives the following recurrence relations
		\begin{eqnarray}
 C_{n+1,l} \sqrt{(n+1)[ k -  (n+l)]}  =  C_{n,l} \label{rel-rec-1}, \quad n = 0, 1, 2, \cdots k-l-1,
		\label{rel-rec-1}
		\end{eqnarray}
with  the following conditions that must be satisfied by the eigenvalues $\lambda_1$ and $\lambda_2$
		\begin{eqnarray}
\lambda_1^{k-l+1} \lambda_2^{l} = 0,
		\label{cond-1}
		\end{eqnarray}
for $l = 0, 1, \ldots, k+1$. Similarly using the action of the operator $j_2^-$ in Eq.(\ref{action2+}), the eigenvalues equation (\ref{Eqlambda}) for  the second mode gives
		\begin{eqnarray}
 C_{n,l+1} \sqrt{(l+1)[ k -  (n+l)]}  =  C_{n,l} \label{rel-rec-1}, \quad l = 0, 1, 2, \cdots k.
		\label{rel-rec-2}
		\end{eqnarray}
The  conditions (\ref{cond-1}) are satisfied simultaneously for the variables $\lambda_1$ and $\lambda_2$ given by
$$ \lambda_1 = \eta z_1, \qquad  \lambda_2 = \eta z_2, \qquad (z_1,z_2) \in \mathbb{C}^2$$
in terms of the generalized Grassmann variable $\eta$ of order $k+1$ $(\eta^{k+1}=0)$.
From the recurrence relation (\ref{rel-rec-1}), one shows
\begin{eqnarray}
 C_{n,l}  = C_{0,l}~ \sqrt{ \frac{(k-n-l)!}{(k-l)!~ n!}}.
\label{sol-rel-rec-1}
\end{eqnarray}
On the other hand, using the recurrence relation (\ref{rel-rec-2}), it is simple to verify that
\begin{eqnarray}
 C_{0,l} = C_{0,0}~ \sqrt{ \frac{(k-l)!}{k!~ l!}}.
\label{sol-rel-rec-2}
\end{eqnarray}
It follows that the expansion coefficients $ C_{n,l} $ are given by
\begin{eqnarray}
 C_{n,l} = C_{0,0} \sqrt{ \frac{(k-n-l)!}{k!~ n! ~ l!}}.
\label{sol-rel-rec-1}
\end{eqnarray}
Finally the eigenstates $\vert \lambda_1 , \lambda_2 \rangle \equiv \vert \eta , z_1, z_2 \rangle $ write
\begin{eqnarray}\label{su3-cs}
\vert \eta , z_1, z_2 \rangle   = {\cal N}_3 \sum_{l=0}^{k} \sum_{n=0}^{k-l} \sqrt{ \frac{(k-n-l)!}{k! ~n!~ l!}}  \eta^{n+l} {z_1}^{n}  {z_2}^{l} \vert n,l \rangle.
\end{eqnarray}
where ${\cal N}_3$ is  fixed from the normalization condition of the states $\vert \theta_1 , \theta_2 \rangle $. Indeed, one gets
$$ \vert {\cal N}_3\vert^{-2} = \sum_{l=0}^{k} \sum_{n=0}^{k-l} \sqrt{ \frac{(k-n-l)!}{k! ~n!~ l!}}  \eta^{n+l} \bar\eta^{n+l} {\vert z_1\vert}^{2n}  {\vert z_2\vert}^{2l}.  $$
The coherent states $\vert \eta , z_1, z_2 \rangle $  satisfy the identity resolution
	\begin{equation}
	\int \vert \eta, z_1 , z_2 \rangle d\mu (\eta ,\bar \eta, z_1, \bar z_1, z_2 , \bar z_2)
	\langle \eta, z_1 , z_2  \vert = \sum_{l = 0}^{k}  \sum_{n = 0}^{k-l}\vert n,l \rangle \langle n,l \vert
	\end{equation}
where the measure $d\mu (\eta ,\bar \eta, z_1, \bar z_1, z_2 , \bar z_2)$ is obtained as for $su(2)$ case. It can written as
$$d\mu (\eta ,\bar \eta, z_1, \bar z_1, z_2 , \bar z_2) = \vert {\cal N}_3 \vert^{-2} \sigma(\eta, \bar \eta) d\eta d\bar\eta~ e^{-(\vert z_1 \vert^2 + \vert z_2 \vert^2)}~\frac{d^2z_1d^2z_2}{\pi^2}.$$
The sum  over the complex variables implies that the function  $\sigma(\eta, \bar \eta) $ must satisfy the integral equation
$$ \int \sigma(\eta, \bar \eta) d\eta d\bar\eta  \eta^{n+l} \bar\eta^{n+l} = \frac{k!}{(k-n-l)!},$$
which  coincides with (\ref{condition mesure2}) for $k = 2j$. Indeed,  one gets
	\begin{equation}\label{sigma}
\sigma(\eta, \bar \eta) = \sum_{n=0}^{k} \frac{1}{k!(k-n)!}\eta^{k-n} \bar\eta^{k-n}.
\end{equation}

\subsection{The  Barut-Girardello coherent states for $su(r+1)$ algebra}
The algebra $su(r+1)$ is defined by the generators $e_i$, $f_i$, $h_i$ ($i = 1, 2,\dots, r$) and the relations
\begin{equation}
[ e_i , f_j ] = \delta_{ij} h_j, \quad
[ h_i , e_j ] = a_{ij} e_j , \quad [ h_i , f_j ] = - a_{ij} f_j
\end{equation}
\begin{equation}
[ e_i , e_j ] = 0 \quad [ f_i , f_j ] = 0 \quad \textrm{for} \qquad \vert i - j \vert > 1
\end{equation}
\begin{equation}
e_i^2e_{i \pm 1} - 2 e_ie_{i \pm 1}e_i + e_{i \pm 1}e_i^2 = 0,\quad
f_i^2f_{i \pm 1} - 2 f_if_{i \pm 1}f_i + f_{i \pm 1}f_i^2 = 0
\end{equation}
where $(a_{ij})_{i,j=1,2,\dots,r}$ is the Cartan matrix of $su(r+1)$, i.e. $a_{ii} = 2$, $a_{i,i\pm 1} = -1$
and $a_{ij} = 0$ for
$\vert i-j \vert > 1$. To construct the Barut--Girardello coherent states of $su(r+1)$ algebras,  we first recall some necessary elements related  to irreducible unitary
representations  for $SU(r+1)$. We denote ${\cal D}^k_r$ as a representation of $SU(r+1)$ where  $k$ determines
the dimension of the representation. The matrix representation is easily obtained by  employing  the bosonic realization in which an adapted basis is given in term of
  $(r+1)$ bosonic pairs of creation and annihilation operators; They satisfy the commutation relations
\begin{equation}
[ b_k^{-} , b_l^{+}] = \delta_{kl},
\end{equation}
where $k,l = 0, 1, 2, .., r$. The simultaneous eigenstates $\vert \vert n_0, n_1,\cdots, n_{r}\rangle\rangle~ (n_i \in \mathbb{N})$ are the tensorial product of the eigenstates  of the
occupation numbers $ b_k^+b_k^-$. The Fock space is then  generated by the states
\begin{equation}
\vert\vert n_0, n_1,\dots, n_{r}\rangle \rangle = \frac{(b_0^+)^{n_0}}{\sqrt{n_0!}}\frac{(b_1^+)^{n_1}}{\sqrt{n_1!}}\cdots
\frac{(b_{r}^+)^{n_{r}}}{\sqrt{n_{r}!}}
\vert\vert 0, 0,\cdots , 0\rangle\rangle.
\end{equation}
In this this bosonic representation, the Weyl generators $e_i$ , $f_i$ and the Cartan operators $h_i$ of the algebra $su(r+1)$  are realized  as follows
\begin{equation}
e_i = b_{i-1}^+ b_{i}^- ,{\hskip 0.5cm} f_i = b_{i-1}^- b_{i}^+ ,{\hskip 0.5cm} h_i = b_{i-1}^+ b_{i-1}^- - b_{i}^+ b_{i}^-,\quad i = 1, 2,\dots, r.
\end{equation}
The irreducible unitary
representation ${\cal D}^k_r$ is obtained within this realization by considering the  subspace
 $$ {\cal E}^k_r = \{ \vert\vert n_0, n_1,\dots, n_{r}\rangle\rangle;  n_0 + n_1 +\dots+ n_{r} = k\}.$$
 of dimension $\frac{(k+r)!}{k!r!}$.  In this symmetric representation the state
of highest weight is $\vert\vert k, 0,\dots, 0\rangle\rangle$.\\

For the Lie algebra $su(r+1)$, the  generators defined in terms of the Weyl operators $f_i$ as
\begin{equation}\label{jiplus}
j^+_1 \equiv f_1, {\hskip 0.5cm} j^+_i \equiv [ f_{i} , j^+_{i-1}] , \quad i = 2,3,\dots, r,
\end{equation}
are mutually commuting.  Similarly, the generators  $j^-_i = (j^+_i)^{\dagger}$ ($i = 1, 2,\dots, r$) defined as
\begin{equation}\label{jimoins}
j^-_1 \equiv f_1, {\hskip 0.5cm} j^-_i \equiv [ j^-_{i-1}, e_{i} ] , \quad i = 2,3,\dots, r,
\end{equation}
commute. The bosonic realization of operators $j^+_i$ (\ref{jiplus}) and $j^-_i$ (\ref{jimoins}) writes as
$$ j^+_i =  b_0^- b_{i}^+ \quad j^-_i = b_0^+ b_{i}^-, \quad i = 1, 2, 3,\dots, r, $$
in terms of the oscillators $b_{i}^-$ and $b_{i}^+$. Using the condition $k = n_0 + n_1 +\dots+ n_{r}$,
one identifies  $\vert\vert n_0, n_1,\dots, n_{r}\rangle\rangle =  \vert\vert k - (n_1 +\dots+ n_{r}) , n_1, n_2, \dots, n_{r}\rangle\rangle \equiv \vert  n_1, n_2\dots, n_{r}\rangle  $ so that the representation space is given by
$$ {\cal E}^k_r = \{ \vert  n_1, n_2,\dots, n_{r}\rangle; n_1+ n_2 +\dots+ n_{r} \leq k\}.$$
Using the usual action of the lowering and raising harmonic oscillator operators  on the Fock number states, one gets
\begin{equation}\label{actionFi}
 j^+_i \vert n_1, n_2,\dots, n_i,\dots, n_{r+1}\rangle =  \sqrt{(n_i+1)(k - (n_1+n_2+\cdots+n_{r}))} \vert n_1, n_2,\dots,n_i+1 ,\dots,n_{r+1}\rangle
\end{equation}
\begin{equation}\label{actionEi}
j^-_i\vert n_1, n_2,\dots, n_i,\dots, n_{r+1}\rangle =  \sqrt{n_i(k +1 - (n_1+n_2+\cdots+n_{r}))} \vert n_1, n_2,\dots,n_i-1 ,\dots,n_{r+1}\rangle.
\end{equation}
We note that the operators $j^+_i$ and $j^-_i$  satisfy the following nilpotency relations
\begin{equation}\label{nilptencyFi}
 (j^+_1)^{n_{1}} (j^+_2)^{n_{2}} \cdots (j^+_r)^{n_{r}} = 0, \quad    (j^-_1)^{n_{1}} (j^-_2)^{n_{2}} \cdots (j^-_r)^{n_{r}} =  0, \quad {\rm for}~~ n_1+n_2+\cdots+ n_{r} = k+1.
\end{equation}
To get the explicit expressions of  the Barut--Giraredello coherent states associated with $su(r+1)$, one has to determine the
common eigenvectors of the ladder generators $j^-_i$ $(i = 1, 2, \cdots, r)$. In this respect, we consider the eigenvalues equations
\begin{eqnarray}\label{Eqsur}
j^-_i \vert \lambda_1 , \lambda_2, \cdots, \lambda_i \cdots, \lambda_{r} \rangle =  \lambda_i \vert \lambda_1 ,\lambda_2, \cdots, \lambda_i \cdots, \lambda_{r} \rangle.
\end{eqnarray}
To find the corresponding solutions, we expand the state $\vert \lambda_1 ,\lambda_2, \cdots, \lambda_i \cdots, \lambda_{r} \rangle $ as follows
		\begin{eqnarray}
\vert \lambda_1 ,\lambda_2, \cdots, \lambda_i \cdots, \theta_{r} \rangle  = \sum_{n_1=0}^{k} \sum_{n_2=0}^{k-n_1} \cdots \sum_{n_{r}=0}^{k-(n_1+\cdots+n_{r-1})} C_{n_1,n_2\cdots,n_{r}}
\lambda_1^{n_1} \lambda_2^{n_2} \cdots \lambda_{r}^{n_{r}}  \vert n_1, n_2,\dots, n_i,\dots, n_{r}\rangle.
		\label{cs-sur}
		\end{eqnarray}
Using the nilpotency conditions (\ref{nilptencyFi}), it is simple to see from the eigenvalues equations (\ref{Eqsur}) that the variables $\lambda_i ~(i=1,2,\cdots,r)$
 satisfy the conditions
	\begin{eqnarray}
\lambda_{1}^{n_{1}} \lambda_{2}^{n_{2}} \cdots \lambda_{r}^{n_{r}} = 0 \quad {\rm for}~~ n_1+n_2+\cdots+ n_{r} = k+1.
\label{nilp-lambda-sur}
		\end{eqnarray}
As discussed previously for $r=1$ and $r=2$, we factorize  the variables $\lambda_i$ ($i = 1, 2, \cdots , r$)
$$\theta_i = \eta z_i$$
in terms of one generalized Grassmann variable $\eta$ and the complex variable $z_i$ so that  the conditions (\ref{nilp-lambda-sur}) are fulfilled.
Reporting the expression (\ref{cs-sur}) in the eigenvalue equation (\ref{Eqsur}) for the mode $i$ $(i=1,2,\cdots,r)$, we obtain the recurrence relation
$$ C_{n_1,n_2, \cdots, n_i-1,\cdots ,n_{r}} = \sqrt{n_i(k- (n_1+n_2+\cdots+ n_i+\cdots+n_{r}))} C_{n_1, n_2, \cdots, n_i,\cdots ,n_{r}}, $$
which gives
$$ C_{n_1,n_2, \cdots, n_i,\cdots ,n_{r}} = \sqrt{\frac{(k - (n_1+n_2+\cdots+ n_i+\cdots+n_{r}))!}{n_i!(k - (n_1+n_2+\cdots+ n_i+\cdots+n_{r})+n_i)!}} C_{n_1,n_2, \cdots, 0,\cdots ,n_{r}}$$
for $n_i = 0, 1, \cdots,  k-( n_1+n_2+\cdots+n_{i-1})$ when $i = 1, 2,  \cdots, r$  with $n_{-1}\equiv 0$.
Using this result, the
explicit form of the expansion coefficients $C_{n_1,n_2, \cdots, n_i,\cdots ,n_{r}}$ is given by
\begin{eqnarray}\label{Cn-sur}
C_{n_1,n_2, \cdots, n_i,\cdots ,n_{r}} = \sqrt{\frac{(k - (n_1+ n_2+\cdots+n_{r}))!}{n_1!n_2!\cdots n_{r}! k!}}  C_{0,0, \cdots ,0},
\end{eqnarray}
where the factor $C_{0,0, \cdots ,0}$ is determined from the normalization condition of the eigenstates
$\vert \lambda_1, \lambda_2 ,  \cdots, \lambda_{r} \rangle \equiv \vert \eta, z_1, z_2 ,  \cdots, z_{r} \rangle $. This gives
		\begin{eqnarray}\label{sur-cs}
\vert \eta, z_1, z_2 ,  \cdots, z_{r} \rangle  = {\cal N}_r\sum_{n_1=0}^{k} \sum_{n_2=0}^{k-n_1} \cdots \sum_{n_{r}=0}^{k-(n_1+\cdots+n_{r-1})}
 \sqrt{\frac{(k - (n_1+n_2+\cdots+n_{r}))!}{n_1!n_2!\cdots n_{r}!k!}}\times\\
\times \eta^{n_1+n_2+\cdots+n_{r}}  z_1^{n_1} z_2^{n_2} \cdots z_{r}^{n_{r}}  \vert  n_1, n_2,\dots, n_{r}\rangle.
		\end{eqnarray}
where the normalization factor writes
$$ \vert {\cal N}_r\vert^{-2}= \sum_{n_1=0}^{k} \sum_{n_2=0}^{k-n_1} \cdots \sum_{n_{r}=0}^{k-(n_1+\cdots+n_{r-1})} \frac{(k - (n_1+n_2+\cdots+n_{r}))!}{n_1!n_2!\cdots n_{r}!k!}
  \vert \eta \vert ^{2(n_1+n_2+\cdots+n_{r})}
 \vert z_1\vert^{2n_1}  \vert z_2\vert^{2n_2} \cdots  \vert z_r\vert^{2n_r}.$$
The vectors $\vert \eta, z_1, z_2 ,  \cdots, z_{r} \rangle $  constitutes an over-complete set of states labelled continuously by
one Grassmann variable $\eta$ and $r$ complex (bosonic) variables $z_i$. The computation of the  measure, with respect which the identity resolution is ensured,
 works like the $su(3)$ case discussed in the previous section. Indeed, we have
\begin{equation}	\int  \vert \eta, z_1, z_2 ,  \cdots, z_{r}  \rangle d\mu_r (\eta ,\bar \eta, \{z_i\}, \{\bar z_i\})
	\langle  \eta, z_1, z_2 ,  \cdots, z_{r}  \vert = {\rm Identity}.
	\end{equation}
The measure $d\mu_r (\eta ,\bar \eta, \{z_i\}, \{\bar z_i\})$,  is given by
$$d\mu (\eta ,\bar \eta, \{z_i\}, \{\bar z_i\}) = \vert {\cal N}_r \vert^{-2} \sigma(\eta, \bar \eta) d\eta d\bar\eta~ e^{- \sum_{i=1}^{r}\vert z_i \vert^2}~\frac{d^2z_1d^2z_2 \cdots d^2z_r}{\pi^r},$$
where the function $\sigma(\eta, \bar \eta)$ is given by (\ref{sigma}). The coherent states (\ref{sur-cs}) reduces to $su(2)$ and $su(3)$ Barut--Girardello coherent states
given by (\ref{BG-Cs}) and (\ref{su3-cs}) for $r=1$ and $r=2$ respectively.

\section{Closing remarks}

The Barut--Girardello coherent states, labelled by complex variables, have been initially proposed
by Barut and Girardello as the eigenstates of the lowering
generators in the representation spaces of Lie algebras \cite{Barut}. This procedure
is only possible for non-compact Lie Groups, as
for instance $SU(1,1)$. Most of the works presented in the
literature on the Barut-Girardello coherent states were done for
 infinite-dimensional representation
algebras.
Recently, the Barut--Girardello
coherent states
for the Lie algebra  $su(2)$ algebra, the Lie algebra $su(1,1)$  with truncated representation space  and  Pegg--Barnett
oscillator algebra were developed in  \cite{DK} using
the formalism of generalized Grassmann
variables \cite{Majid}.\\

In this paper we developed further the idea proposed   in \cite{DK}. We define first
the generalized Weyl-Heisenberg algebra by mean of $r$ pairs of
creation annihilation operators and the associated Fock--Hilbert space. This algebra extends the
generalized oscillator algebras introduced in \cite{daoud-kibler2} for $r=1$ and in \cite{daoud-kibler3} for $r=1$.
The Hilbertian and analytical representation were investigated. In the Fock-Bargmann
representation, we employed the generalized Grassmann variables. A special attention were devoted
to the definition of this kind of exotic variables in terms of the ordinary Grassmann commonly
used in the formulation of super-symmetric models. This connection provides us with
the appropriate scheme to define the derivative
and integration operations by exploiting  the properties of the ordinary  Grassmann-valued variables.  The
essential observation  arising  from this realization is the algebraic description of qukits in terms
of a symmetric ensemble of qubits. In this sense, the generalized
Weyl-Heisenberg algebras provide the algebraic framework to describe qubits and qukits which are
the basic ingredients in quantum information theory. Furthermore, the method, based on generalized Weyl--Heisenberg algebras, that we adopted in this work  is an alternative way
to define   the generalized
Grassmann variables without resorting to the analysis developed in the context of quantum algebras (see \cite{DK} and references therein).
This simplifies the
 construction of  the coherent states \`a la Barut--Girardello
associated with the $su(r+1)$ Lie algebra and their over--completion properties. We hope that this novel
construction of generalized Grassmann variables for qukit systems
and Barut-Girardello coherent states for $su(r+1)$ algebras will
be of interest in the field of quantum
systems with finite dimensional Hilbert space , especially
for  pseudo-Hermitian quantum systems. Also, it is interesting to investigate the relation between
the generalized Weyl--Heisenberg algebras discussed in this paper and the formalism of nonlinear fermions discussed
in \cite{Trifonov1}.\\


{\bf Acknowledgments:}
M.D. would like to thank ICTP for hospitality during the completion of this work,
which was done in the frame of the Associate Membership Programme of the ICTP.
L. G. is supported by the Abdus Salam International
Centre for Theoretical Physics (ICTP).

\end{document}